\documentclass[]{aastex}
\usepackage{emulateapj5}
\begin{document}

\title{An Empirical Decomposition of Near--IR Emission into 
Galactic and Extragalactic Components}
\shorttitle{Empirical Decomposition of Near--IR Emission}

\author{Richard G. Arendt\altaffilmark{1} and Eli Dwek}
\affil{NASA's Goddard Space Flight Center}
\affil{Code 685, Greenbelt, MD, 20771}
\email{arendt@stars.gsfc.nasa.gov, edwek@stars.gsfc.nasa.gov}
\altaffiltext{1}{Science Systems \& Applications, Inc.}

\begin{abstract}
We decompose the {\it COBE}/DIRBE observations of the near--IR sky 
brightness (minus zodiacal light) into 
Galactic stellar and interstellar medium (ISM) components and an extragalactic background. 
This empirical procedure allows us to estimate the 4.9 $\micron$ cosmic infrared 
background (CIB) as a function of the CIB intensity at shorter wavelengths. 
A weak indication of a rising CIB intensity at $\lambda > 3.5$ $\micron$ hints
at interesting astrophysics in the CIB spectrum, or warns 
that the foreground zodiacal emission may 
be incompletely subtracted. Subtraction of only the stellar component from the 
zodiacal--light--subtracted all--sky map reveals the 
clearest 3.5 $\micron$ ISM emission map, which is found to be
tightly correlated with the ISM emission at far--IR wavelengths.
\end{abstract}


\section{Introduction}

The principal scientific goal of the Diffuse Infrared Background Experiment
(DIRBE; Silverberg et al. 1993) aboard the {\it Cosmic Background Explorer} ({\it COBE}; 
Boggess et al. 1992) spacecraft was
to detect the cosmic infrared background (CIB) in 10 broad bands from the 
near--infrared (near--IR) at 1.25 $\micron$ to the far--IR at 240 $\micron$. The energy in the 
CIB arises from thermonuclear processes within stars, and to a lesser 
extent from gravitational energy release by active galactic nuclei (AGN). At
near--IR wavelengths, these energy sources are observed directly in the CIB.
At the far--IR wavelengths, the CIB is thermal emission from dust that is heated
by either starlight or the emission from AGNs. At all IR wavelengths, the 
CIB will contain a blend of emission from galaxies at all redshifts. 
As such, the CIB forms an integral constraint for models of the dominant 
processes of energy release over the history of the universe.

The DIRBE succeeded at detecting the CIB at 140 and 240 
$\micron$ (see Hauser \& Dwek 2001 for a review).
In these bands, local foregrounds from interplanetary dust (i.e. zodiacal light)
and our own Galaxy's interstellar medium (ISM) are only about as bright as the CIB itself. At the 
shorter wavelengths, Galactic stars and zodiacal emission are brighter than the CIB. 
The latter foreground can be removed by modeling its spatial and temporal 
variations over the course of the DIRBE mission. Most of the uncertainties in 
the removal of the zodiacal light arise from uncertainties in the intensity of its 
isotropic component. The main uncertainties in the removal of the Galactic 
stellar emission arise from the removal of the diffuse emission from faint 
unresolved stars, since bright resolved stars can be readily removed from 
the DIRBE maps by blanking. Various methods have been used to remove 
the diffuse Galactic stellar emission component: (1) Hauser et al. (1998) 
used a faint star model (Arendt et al. 1998), based on SKY, a star count 
model developed by Wainscoat et al. (1992). Uncertainties associated 
with this stellar emission model and the removal of zodiacal light prevented 
the detection of the CIB at near--IR wavelengths; (2) Dwek \& Arendt 
(1998) created an empirical map of the starlight at 2.2 $\micron$ directly 
from the DIRBE map, and used it as a spatial template for the starlight emission 
at the other near--IR wavelengths. Their procedure 
yielded a 3.5 $\micron$ CIB estimate that was definitely non--zero, but did 
not meet strict isotropy requirements; (3) Wright \& Reese (2000) used a 
statistical method, comparing histograms of simulated star counts with 
histograms of map intensities, to determine the residual background 
emission at 2.2 and 3.5 $\micron$; 
(4) Gorjian, Wright \& Chary (2000) subtracted the contribution of stars resolved by 
ground--based observations at 2.2 and 3.5 $\micron$ from a dark patch of the 
DIRBE sky to estimate the CIB at these wavelengths. Wright (2001), 
and Cambr\'esy et al. (2001) 
performed a similar analysis at 1.25 and 2.2 $\micron$ using
Two Micron All Sky Survey (2MASS) star counts to subtract the bulk of the Galactic 
stellar emission. Each of these studies arrived at an estimate of the CIB at these 
wavelengths; (5) A very different approach to detecting the CIB has been pursued by 
Kashlinsky et al. (1996a, 1996b) and Kashlinsky \& Odenwald (2000). 
In these studies, the random fluctuations of the confusion limited CIB 
is measured instead of the mean level. Various measures of these fluctuations
(e.g. variance, two--point correlation function, power spectrum) can be related 
to the mean CIB in the context of a chosen cosmological model.
This approach yields results that are independent of the uncertainties 
in the zero-point of the DIRBE data and the foreground emission models,
but do rely on accurate assessment and modeling of the instrument and 
foreground variations.

The positive detection of the near--IR CIB with the Near Infrared Spectrometer (NIRS) on
the {\it Infrared Telescope in Space} ({\it IRTS}) by Matsumoto et al. (2000) is very 
important, since it is the only measurement of the absolute CIB intensity 
based on a data set other than the DIRBE data. 
However, while the raw data sets are independent measurements, 
this analysis of the NIRS data did use the Kelsall et al. (1998) model to subtract 
the zodiacal light emission from the NIRS sky maps. 

In this paper, we adopt an empirical approach to modeling the Galactic 
stellar and ISM emission at 3.5 $\micron$. Observations at another wavelength 
are used as an emission template which is than scaled by a constant
multiplicative factor to represent the 3.5 $\micron$ emission. Dwek \& Arendt (1998) 
used this approach to model stellar emission at observed by DIRBE at 
1.25, 3.5, and 4.9 $\micron$ using the 2.2 $\micron$ emission as a one--component 
model template. Puget et al. (1996) and Schlegel, Finkbeiner, \& Davis (1998)
have used this approach in modeling zodiacal emission at wavelengths $> 100$ 
$\micron$ with a scaled version of the 25 $\micron$ emission. Arendt et al. (1998) 
modeled ISM emission in this manner using the 100 $\micron$ emission as a spatial 
template at 3.5 -- 240 $\micron$, and, in turn, 21 cm H I emission as a template 
for 100 $\micron$ emission. The common drawback of all these 
models is that with only a single spatial component, they cannot account for any 
deviation from the mean spectrum (or scaling factor) of the modeled emission. 
Arendt et al. (1998) addressed the problem in the context of the 240 $\micron$ 
ISM emission by constructing a two--component model using both the 100 and 140 
$\micron$ emission as templates. Observationally, this was suggested by the fact 
that the color--color plot at these wavelengths exhibits a fairly tight correlation, which 
is associated with variation of the mean temperature of ISM dust grains.
Functionally, the use of two templates provides an estimate of both the intensity 
and the color of the emission at all locations. This enables a better extrapolation of the 
observed emission to other wavelengths. Equivalently, a two--component model can 
be viewed as a means of decomposing emission that arises from two spatially 
and spectrally distinct phases or populations. Such a decomposition can be applied 
even if the templates themselves are a combination of the true physical 
components, as long as they are not combinations in the same proportions. 
This two-component ISM model was a significantly more accurate representation 
of the ISM emission, which allowed Hauser et al. (1998) to demonstrate isotropy of the 
detected CIB emission. Following this example, we now develop an empirical 
two--component model for the near-IR Galactic stellar emission. One version of this 
model will ignore extinction effects, and is used only at higher Galactic latitudes. 
A second version of the model includes adjustments for extinction, and is derived and 
applied at all latitudes. Without the adjustments, the color variations traced by the 
two component model would be caused by interstellar reddening, rather than by 
intrinsic changes in the stellar emission. For the ISM emission, which is much fainter than 
the stellar emission in the near--IR, we continue to use the one--component model of
Arendt et al. (1998), although here the parameters of the ISM and stellar models will 
be derived simultaneously. 

In \S2 we present our new model for the 3.5 $\micron$ zodiacal-light-subtracted
emission in which the stellar component is represented by a linear combination of the 
extinction--corrected stellar emission at 2.2 and 4.9 $\micron$, the ISM 
component is modeled using its far--IR emission, and the CIB is assumed
to be isotropic. 
The optical depths at these wavelengths are expressed in terms of the 
stellar emission at 1.25 and 2.2 $\micron$, where the extinction is largest 
and the dust emission negligible. The Rieke \& Lebofsky (1985) extinction 
law is used to scale the extinction to other wavelengths. The residual 
emission, obtained after subtraction of the stellar and ISM emission components 
from the zodi--subtracted 3.5 $\micron$ DIRBE maps, should be equal to 
the 3.5 $\micron$ CIB. In practice, we show that with this purely empirical 
model the 3.5 $\micron$ residual is a linear combination 
of the CIB emission at 2.2, 3.5, 4.9, and 100 $\micron$. 
In \S3 we report the level and uncertainties of the residual emission, and discuss 
its isotropy in \S4. These results, combined with those of previous studies, allow
us to constrain the CIB at 4.9 $\micron$ (\S5). In \S6, we show that our model 
leads to a clearer map of the large--scale near--IR ISM emission than has previously 
been obtained. The results of our paper are briefly summarized in \S7.

\section{Decomposition of the 3.5 \lowercase{$\mu {\rm m}$} Emission}

\subsection{Modeling Galactic Stellar, ISM, and Extragalactic Components}

The observed 3.5 $\micron$ emission [after subtraction 
of the zodiacal light emission by Kelsall et al. (1998)] is the sum of 
Galactic stellar and ISM foreground emission components and a 
cosmic infrared background
\begin{equation}
I_{obs}(3.5) = I_{star}(3.5) + I_{ISM}(3.5) + I_{CIB}(3.5).
\end{equation}

Dwek \& Arendt (1998) modeled the 3.5 $\micron$ Galactic stellar emission,
$I_{star}(3.5)$, 
by using the observed 2.2 $\micron$ emission as an empirical spatial 
template with a scaling factor determined by a least--squares fit. 
In the present analysis we use the same general approach of employing 
the observed emission at other wavelengths as empirical spatial 
templates of the  3.5 $\micron$ Galactic stellar emission, but we make
modifications in two respects. First, we use {\it two} spatial templates, the 2.2
$\micron$ emission and the 4.9 $\micron$ emission. This allows a more
accurate modeling of the 3.5 $\micron$ emission assuming that there
are two distinct components to the stellar emission at each of these wavelengths.
In this case, the linear combination of the components that produces the 3.5
$\micron$ emission can be expressed as a (different) linear combination 
of the 2.2 and 4.9 $\micron$ emission. 
The second change from the Dwek \& Arendt (1998) procedure is that here
we apply adjustments to the 2.2 and 4.9 $\micron$ spatial templates to 
account for extinction differences between those wavelengths and 
3.5 $\micron$. This correction is intended to make the spatial templates 
(especially at 2.2 $\micron$) more accurate at low latitudes ($|b| < 10\arcdeg$). 
The extinction corrections involve the simplifying assumptions 
that extinction can be treated as a foreground screen, and
that the intrinsic background sources have uniform color. 
Arendt et al. (1994) have previously demonstrated that the near-IR colors 
of the integrated starlight measured by DIRBE at $|b| < 5\arcdeg$ can be 
fit very well making these assumptions. At high latitudes the 
extinction is so low at these wavelengths that the 
correction has no significant effect. 

Thus, we represent the stellar emission at 
3.5 $\micron$ as a linear combination of the 
extinction--corrected emission at 2.2 and 4.9 $\micron$:
\begin{equation}
I_{star}(3.5) = [A I_{star}(2.2)e^{\tau_{2.2}} + 
                B I_{star}(4.9)e^{\tau_{4.9}}] e^{-\tau_{3.5}}.
\end{equation}
where $I_{star}(\lambda)$ and $\tau_{\lambda}$ are, respectively, the 
stellar component of the observed emission
and the optical depth at the wavelength $\lambda$. 
With the use of an adopted reddening law to define the wavelength dependence of the 
extinction, the extinctions in eq.~(2) can be expressed in terms of the ratio
of the 1.25 to 2.2 $\micron$ intrinsic stellar intensity:
\begin{equation}
e^{-(\tau_{1.25}-\tau_{2.2})} \propto \frac{I_{star}(1.25)}{I_{star}(2.2)}.
\end{equation}
This ratio of the two shortest wavelength DIRBE bands is used because extinction 
will be strongest, and ISM emission appears to be negligible in these bands
(Arendt et al. 1994; Freudenreich 1996).
Thus, the actual model for the stellar emission component is given by 
\begin{eqnarray}
I_{star}(3.5) &  = &  
A I_{star}(2.2) \left[\frac{I_{star}(1.25)}{I_{star}(2.2)}\right]^{\alpha} \nonumber \\
&  & + B I_{star}(4.9) \left[\frac{I_{star}(1.25)}{I_{star}(2.2)}\right]^{\beta}
\end{eqnarray}
where $\alpha = (\tau_{3.5}-\tau_{2.2}) / (\tau_{1.25}-\tau_{2.2})$ 
and $\beta = (\tau_{3.5}-\tau_{4.9}) / (\tau_{1.25}-\tau_{2.2})$. The
numerical values of $\alpha = -0.317$ and $\beta = 0.210$ are calculated 
using the Rieke \& Lebofsky (1985) reddening law, which has been 
shown to be a good representation of the extinction observed by DIRBE 
(Arendt et al. 1994).

Emission from the ISM, the second term in eq.~(1), is relatively faint at near--IR 
wavelengths, but it is not negligible at $\lambda \gtrsim 3.5 \micron$. 
Our model of the ISM emission is the same simple 
one-component model used by Hauser et al. (1998):
\begin{equation}
I_{ISM}(3.5) = C I_{ISM}(100)
\end{equation}
where $ I_{ISM}(100)$ is the 100 $\micron$ ISM intensity.
This model is obviously a valid approximation at low optical depth,
but remarkably this linear correlation is still observed to exist at 
very low latitudes where high optical depths might be expected. 
This is demonstrated and discussed in \S 6.

The third component of the observed emission, the CIB, is modeled as an 
isotropic constant term. While the Galactic stellar and ISM emission components are 
modeled using their spatial structure, it is the {\it lack} of structure that distinguishes the
CIB. Even though a few of the nearest, brightest galaxies are resolved by DIRBE's
$0\fdg7$ beam (Odenwald, Newmark, \& Smoot 1998), we expect 
that because of stellar confusion 
and instrumental noise, any detection of the CIB will be isotropic to first order.

Thus, the complete model of the observed 3.5 $\micron$ emission is represented by 
\begin{eqnarray}
I_{obs}(3.5) & = &
A I_{star}(2.2)\left[\frac{I_{star}(1.25)}{I_{star}(2.2)}\right]^{\alpha}\nonumber\\
& & + B I_{star}(4.9)\left[\frac{I_{star}(1.25)}{I_{star}(2.2)}\right]^{\beta}\nonumber\\
& & + C I_{ISM}(100) + I_{CIB}(3.5).
\end{eqnarray}
In the event that the extinction is uniform or negligible, the extinction 
terms can be rolled into the $A$ and $B$ 
coefficients, yielding the ``no--extinction'' model:
\begin{eqnarray}
I_{obs}(3.5) & = & A I_{star}(2.2) + B I_{star}(4.9)\nonumber\\
& & + C I_{ISM}(100) + I_{CIB}(3.5).
\end{eqnarray}\label{noext}
In the following, we refer to this simpler, no--extinction model as ``Model 1''
and refer to the full model [eq.~(6)] with extinction as ``Model 2''.

\subsection{Interpretation of the Residual Emission}

If we had templates of purely stellar emission at 1.25, 2.2 and 4.9~$\micron$
and pure ISM emission at 100~$\micron$, then we could plug these into 
eqs.~(6) or (7), and perform a least--squares fit to determine the coefficients
$A$, $B$, $C$, and the CIB $I_{CIB}(3.5)$. However, what is actually available is 
the {\it observed} emission at 1.25, 2.2, 4.9, and 100 $\micron$, which will contain 
additional ISM and CIB components. Thus for Model 1, eq.~(7), 
we evaluate a least--squares fit to 
determine the coefficients $A'$, $B'$, $C'$, and $D'$ in the equation
\begin{equation}
 I_{obs}(3.5) = A'I_{obs}(2.2) + B'I_{obs}(4.9) + C'I_{obs}(100) +D' 
\end{equation}
which can be expanded [using eq.~(1)] as:
\begin{eqnarray}
I_{obs}(3.5) &=& A' [I_{star}(2.2) + I_{ISM}(2.2) + I_{CIB}(2.2)] \nonumber\\
& & + B' [I_{star}(4.9) + I_{ISM}(4.9) + I_{CIB}(4.9)] \nonumber\\
& & + C' [I_{ISM}(100) + I_{CIB}(100)]\nonumber\\
& & + D'.
\end{eqnarray}
Comparison with eq.~(7) then shows that 
\begin{eqnarray}
A & = & A' \nonumber\\
B & = & B' \nonumber\\
C & = & C' + A' I_{ISM}(2.2) / I_{ISM}(100)\nonumber\\
  &   & + B' I_{ISM}(4.9) / I_{ISM}(100) \nonumber\\
I_{CIB}(3.5)  & = & D' + A' I_{CIB}(2.2) \nonumber\\
  &   & + B' I_{CIB}(4.9) + C' I_{CIB}(100).
\end{eqnarray}
Thus, we see that in solving eq.~(8) for $A'$, $B'$, $C'$, and $D'$, 
instead of solving for $A$, $B$, $C$, and $I_{CIB}(3.5)$
in eq.~(7), we derive only a linear combination of the CIB at 2.2, 3.5,
4.9, and 100 $\micron$, and not the 3.5 $\micron$ CIB directly.

Similarly, when Model 2 [given by eq.~(6)] is evaluated using the {\it observed} 
emission, we start with
\begin{eqnarray}
I_{obs}(3.5) & = & 
A' I_{obs}(2.2)\left[\frac{I_{obs}(1.25)}{I_{obs}(2.2)}\right]^{\alpha}\nonumber\\
& & + B' I_{obs}(4.9)\left[\frac{I_{obs}(1.25)}{I_{obs}(2.2)}\right]^{\beta}\nonumber\\
& & + C' I_{obs}(100) + D'.
\end{eqnarray}
Then, after expansion using eq.~(1) and comparison with the idealized form in 
eq.~(6), we find
\begin{eqnarray}
I_{CIB}(3.5)&=&D' + [A'\alpha r_{12}^{\alpha-1} + 
B'\beta r_{12}^{\beta} r_{41}] I_{CIB}(1.25)\nonumber\\
 & &+[A'(1-\alpha)r_{12}^{\alpha} + B'\beta r_{12}^{\beta} r_{42}] I_{CIB}(2.2)\nonumber\\
 & &+ B' r_{12}^{\beta} I_{CIB}(4.9) + C' I_{CIB}(100)
\end{eqnarray}
after some algebraic reshuffling and keeping only first-order terms.
Here $r_{ij} = I_{star}(\lambda_i)/I_{star}(\lambda_j)$ is the intrinsic 
color of the stellar emission, which is assumed to be constant over the 
regions of sky being examined.

\section{Intensity of the Residual Emission}

The coefficients $A'$, $B'$, $C'$, and $D'$ were determined with a simple
least--squares fit, between the DIRBE zodiacal-light-subtracted maps 
at 1.25 -- 4.9, and 100 $\micron$. These fits minimized the $\chi^2$ 
parameter measuring the goodness of fit between the data, $I_{obs}(3.5)$, and 
the model, right--hand side eqs. (8) or (11). The fit (and all other quantitative 
analysis) was performed using
the approximately equal--area pixels of the native {\it COBE} sky--cube 
projection (Hauser, et al., 1997). 
(Reprojected images in Figs. 1 and 7 are only for display.)
The fit used equal weights for all pixels, except for assigning zero weights to 
the same bright stars
and low latitude emission that were excluded at 3.5 $\micron$ in the 
Hauser et al. (1998) analysis. 
For the no-extinction model, Model 1, we additionally excluded regions at low Galactic 
latitude ($|b| < 10\arcdeg$) in order to prevent the model from 
attempting to match low--latitude color variations at 
the expense of a good fit at high latitudes.
The derived values of the model coefficients and their 1$\sigma$ statistical (random)
uncertainties are listed in Table 1. 
In Figure 1, we show the residual map that results from subtraction of
the stellar and ISM components of each model from the 3.5 $\micron$ emission. 
This residuals have a mean level of $D'$ (in the unblanked regions). They are
depicted on a comparable intensity range as the residuals of Hauser et al. (1998) and 
Dwek \& Arendt (1998), which are also shown in Figure 1. 

For Model 1, the mean intensity of residual emission [eq.~(10)]
becomes:
\begin{eqnarray}
D' & = & I_{CIB}(3.5) - 0.445 I_{CIB}(2.2)\nonumber\\
 &   & - 0.220 I_{CIB}(4.9) -
0.00135 I_{CIB}(100)\nonumber\\
 & = & -0.00197
\end{eqnarray}
where all intensities are in MJy sr$^{-1}$, or 
\begin{eqnarray}
D' & = & \nu I_{CIB}(3.5) - 0.280\ \nu I_{CIB}(2.2)\nonumber\\
 &   & - 0.308\ \nu I_{CIB}(4.9) - 
0.0386\ \nu I_{CIB}(100)\nonumber\\
 & = & -1.69
\end{eqnarray}
with $\nu I_{\nu}$ in nW m$^{-2}$ sr$^{-1}$.
The above expressions show that the
unsubtracted backgrounds in the 2.2, 4.9, and 100 $\micron$ components of the 
models reduce the value of the residual $D'$ from the level of the 3.5 
$\micron$ CIB.

For Model 2, we inserted the numerical
values $r_{12} = 1.2$, $r_{41} = 0.25$, $r_{42} = 0.3$, 
which were determined at midlatitudes where extinction and the CIB 
and ISM emission components should be relatively small 
(e.g. Arendt et al. 1994; Bernard et al. 1994). Then, using the coefficients 
of the fit for this model, we arrive at
\begin{eqnarray}
D' & = & I_{CIB}(3.5) + 0.103 I_{CIB}(1.25) -0.564 I_{CIB}(2.2)\nonumber\\
 &   & 
-0.257 I_{CIB}(4.9) - 0.00123 I_{CIB}(100) \nonumber\\
 & = & -0.00310
\end{eqnarray}
where all intensities are in MJy sr$^{-1}$, or 
\begin{eqnarray}
D' & = & \nu I_{CIB}(3.5) + 0.0368\ \nu I_{CIB}(1.25)\nonumber\\
 &   & -0.355\ \nu I_{CIB}(2.2)\nonumber\\
& & -0.360\ \nu I_{CIB}(4.9) - 0.0351\ \nu I_{CIB}(100)\nonumber\\
 & = & -2.66
\end{eqnarray}
with $\nu I_{\nu}$ in nW m$^{-2}$ sr$^{-1}$.
Not surprisingly, these expressions are similar to the results for Model 1, 
with the addition of a relatively small additive component from the 
1.25 $\micron$ CIB.

As discussed by Hauser et al. (1998), the systematic uncertainties dominate 
the random statistical uncertainties in obtaining the mean 3.5 $\micron$ residual.
Systematic uncertainties include the detector offset and zodiacal light 
uncertainties, and the uncertainty in the stellar and ISM model developed here. 
The zodiacal light uncertainties used by Hauser et al. (1998) are equally 
appropriate here, and are reproduced in Table 2. Detector offset uncertainties
are negligible in comparison, and are therefore not listed. 
The 1~$\sigma$ uncertainties in the 
new stellar and ISM models are estimated from the maximum difference 
of the mean residual intensity of the various high latitude regions 
listed in Table 3 (see below). 
As in Hauser et al. (1998), we take the quadrature sum of the zodiacal light
and combined stellar and ISM uncertainties to arrive at the total systematic
uncertainties, listed for each model in Table 2. The scale factors (weights) that 
are applied to the zodiacal light uncertainties are the same as those used 
to scale each wavelength in the construction of the stellar and ISM models.

\section{Isotropy of the Residual Emission}

While the residual emission, $D'$, derived here is not the CIB directly, it should 
still be isotropic because it is a combination of isotropic backgrounds. 
In fact, $D'$ should be more isotropic than the actual CIB at any single 
wavelength, because any structure in the CIB is likely to be correlated between 
wavelengths and at least partially cancel in the construction of $D'$.
Therefore, following the
example of Hauser et al. (1998), we have examined several different
means of assessing the isotropy of the residual emission. 
In all cases, the key issue is whether or 
not the data show only the amount of variation expected within the 
uncertainties. Isotropy of the residual provides more credibility 
for the accuracy of the subtraction of the foregrounds. It is also a necessary 
condition for attributing the residual to the combination of CIB backgrounds. 

\subsection{Mean Patch Brightness}

The simplest test of isotropy is comparison of the mean brightness in 
selected patches. Table 3 lists the brightnesses derived at several of the
regions examined by Hauser et al. (1998) and Dwek \& Arendt (1998).
The differences in brightness among these patches are smaller than 
in either of the two previous studies, and are smaller than the expected 
systematic uncertainty (even if only the zodiacal light component 
of the uncertainty is considered). The patch intensities for Model 1 are 
not quite as uniform as those for Model 2, but in both cases the differences
are smaller than the expected systematic uncertainties. 

\subsection{Brightness Distributions}

The next isotropy test is the examination of the intensity histograms of the
residual emission. For a truly isotropic residual, the shape and width
of the histograms will reflect only the random noise uncertainties.
Histograms constructed for the north and south portions of the high-latitude 
HQA and HQB regions examined by Hauser et al. (1998) are shown in Figures
2 and 3. The results of fitting Gaussian distributions to the histograms
are shown in the figures and listed in Table 4. The new results exhibit smaller
dispersion than the previous results, particularly over the relatively
large HQA regions. North-south asymmetries in the means of the distributions 
are also reduced in the new results. Kolmogorov-Smirnov (K-S) tests applied
to the distributions, indicate that the HQBN residuals are essentially 
Gaussian for the new results with Model 1 and especially Model 2, 
while distributions in the HQBS (and for the Dwek \& and Arendt (1998) 
HQB regions) are probably only slightly non-Gaussian.
The K-S probabilities of the distributions differing from Gaussian are listed 
in Table 4. The probability is that of finding a random sampling of 
the assumed parent Gaussian distribution with a smaller K-S $D$ parameter 
than the observed distribution. A value of 1.00 is definitely non-Gaussian; a 
value of 0.50 indicates the distribution is indistinguishable from a randomly
sampled Gaussian distribution; a probability of 0.0 would indicate a perfect 
Gaussian distribution with no sampling variation.

\subsection{Systematic Spatial Variations}

The tests above are necessary conditions for isotropy, but they are insensitive
to any anisotropic structure that does not alter the mean level or 
introduce some skewness or tails to the brightness distributions. A simple 
means of looking for likely spatial structure is to check for gradients
in the residual emission. Figure 4 shows the gradients in the residual 
emission as a function of the cosecant of Galactic latitude. Data for the
north and south hemispheres are shown separately. By this test, the Dwek \& 
Arendt (1998) model removed Galactic emission more effectively than the 
models used by Hauser et al. (1998). The present models are seen to make 
further improvements, most dramatically at lower latitudes. The residual 
emission of Model 2 exhibits smaller gradients than that of Model 1 over 
this range of latitudes. Table 5 lists
the gradients for the HQA and HQB regions, showing no significant Galactic 
gradient in the HQB region for the residual emission derived from Model 2. The
derived gradients in the HQA region are even smaller, but statistically 
significant as the region is about ten times larger. Correlation coefficients are 
also listed in Table 5. The correlation coefficient should go to 0.0 as the 
gradient becomes small relative to the variance in the data.

\subsection{Two-Point Correlation Functions}

The most rigorous test of spatial isotropy that was used by Hauser et al. 
(1998) is the two-point correlation function. A truly isotropic residual should
show no structure above the expected random noise on any spatial scales. 
The two-point correlation functions for the 3.5 $\micron$ residual emission 
from the present and previous results are shown in Figure 5. We only examine 
the HQB regions here, since the HQA regions have clearly demonstrated 
anisotropy in the previous tests. Because the Hauser et al. (1998) 
results relied on a statistical model for 
the stellar emission, the mean level but not the detailed structure of
the faint stars was removed from the data. Thus, the Hauser et al. results have
a much higher ``random'' noise level than either the Dwek \& Arendt (1998) or 
the present results. The variance of the residual emission in the present
results is slightly smaller than for the Dwek \& Arendt results.
The two-point correlation function over the HQB region for Model 1 appears 
to be very nearly random by this test. The Model 2 results are not quite as
isotropic as the Model 1 results, but are distinctly better than the Dwek \& 
Arendt results. Detailed examination of the images (Figure 1) suggests that the 
two-point correlation functions in the HQB regions are sensitive to small 
systematic errors in the structure of the zodiacal light model. With the 
two-component models used here, zodiacal light errors at different wavelengths partially 
cancel one another. This cancellation turns out to be slightly more effective
for the parameters of Model 1 than those of Model 2. 

\section{The Cosmic IR Background at 4.9 \lowercase{$\mu {\rm m}$}}

Subtraction of either of the two--component models of stellar emission 
developed here yields a highly isotropic residual emission map. As
shown in \S\S 2.2 and 3, this residual will contain a combination of the CIB at 
several wavelengths, and thus is not a direct measure of the CIB at any one
wavelength. Still, the relations between the CIB intensities defined by 
eqs.~(14) and (16) can be used to place constraints on the 
CIB and its spectrum.

The most straightforward use of the constraints of eqs.~(14) and (16) is to 
apply known or assumed CIB intensities at several wavelengths in order to 
derive the unknown CIB intensity at one particular wavelength. This procedure 
is most useful for estimating the CIB intensity at 4.9 $\micron$, since this
is the only near-IR DIRBE band for which no detection of the CIB has been claimed.
Various estimates of the CIB at 1 -- 3.5 $\micron$ using both {\it COBE}/DIRBE and 
{\it IRTS}/NIRS data are in good agreement, apart from systematic differences
caused by the method chosen for subtraction of zodiacal light
(Hauser et al. 1998; Dwek \& Arendt 1998; Wright 2001; Gorjian et al. 2000; 
Wright \& Reese 2000; Matsumoto et al. 2000; Cambr\'esy et al. 2001; 
see Hauser \& Dwek 2001 for a review). Analyses using the Kelsall et al. (1998)
zodiacal light model derive a brighter 1 -- 3.5 $\micron$ CIB than those
using a version of Wright et al.'s zodiacal model with an additional constraint applied
to the zodiacal light emission (e.g. Wright \& Reese 2001). These systematic 
differences are about as large as the estimated uncertainties in the zodiacal light 
models.
The CIB intensity at 
100 $\micron$ appears to be $\nu I_{CIB}(100) \approx 23 \pm 6$ nW m$^{-2}$ sr$^{-1}$
(Hauser et al. 1998; Lagache et al. 2000; Finkbeiner, Davis, \& Schlegel 2000).
All these prior CIB measurements are listed in Table 6, although not all the
estimates are $3\sigma$ detections or have demonstrated isotropy.
Still, using these CIB values in eq.~(14) or (16) leads to the derived estimates of the 4.9
$\micron$ CIB that are listed in Table 6. 
The 4.9 $\micron$ results do not depend strongly on the 1.25 or 100 $\micron$ backgrounds 
because they only enter through the small extinction correction and the faint
ISM emission respectively. However, the results are sensitive to the 2.2 and 3.5 
$\micron$ CIB. Propagation of the uncertainties indicates that
this is {\it not} a positive detection of the 4.9 $\micron$ CIB, and even the 2--$\sigma$
upper limits on the 4.9 $\micron$ CIB are no better than those already established by 
Hauser et al. (1998) or Dwek \& Arendt (1998): $<41 (24.8 \pm 8)$ and $<36 (23.3 \pm 6.4)$
nW m$^{-2}$ sr$^{-1}$ respectively (values in parentheses are measured residual levels 
and 1--$\sigma$ uncertainties). Figure 6 shows the the 4.9 $\micron$ 
residual that we derive, for assumed CIB intensities at 1.25, 2.2, and 3.5 $\micron$ as 
described above. Regardless of which model is used for zodiacal light subtraction, 
the intensity of the residual emission falls from 1.25 to 3.5 $\micron$, but rises again at 4.9 
$\micron$. 

The apparent rise at 4.9 $\micron$ could indicate thermal dust emission in the CIB spectrum,
or else the stellar emission from a burst of star formation at $z \gtrsim 8$. 
Comparison with galaxy template spectra from Chary \& Elbaz (2001) suggest that if the 
apparent rise at 4.9 $\micron$ is from local luminous galaxies containing hot dust, then the associated 
cold dust emission is likely to exceed the observed far--IR CIB (Fig. 7a, dotted line).
Less luminous galaxies will not exceed the far--IR emission, but lack 
sufficient hot dust to match the near--IR colors of the residual (Fig. 7a, dashed line). 
Extremely red--shifted stellar emission, on the other hand, would pose no 
conflict for the far--IR CIB measurements (Fig. 7a, solid line). 
However, the intensity of 
this emission would be in conflict with the indirect upper limits on the CIB placed by the
TeV $\gamma$--ray measurements and estimates of the CIB fluctuations 
[see Hauser \& Dwek (2001), and references therein]. These upper limits are typically 
5 -- 15 nW m$^{-2}$ sr$^{-1}$ at wavelengths $>5$ $\micron$.
Additionally, the integrated intensity (1--100 $\micron$) of this component would be
 $\sim 57$ nW m$^{-2}$ sr$^{-1}$.
Production of this integrated intensity would require a constant luminosity density of 
${\cal L} \approx 360 \times 10^9$ $L_{\sun}$ Mpc $^{-3}$, or a cosmic star formation rate 
of $25 - 50$ $M_{\sun}$ yr$^{-1}$ Mpc$^{-3}$ at $z>8$ using the relations 
provided by Hauser \& Dwek (2001). This luminosity density and cosmic star formation
rate are high compared to even the most extreme rates recently proposed by Lanzetta 
et al. (2001), and the baryon density of the universe: 
$(0.0192 \pm 0.0018) \rho_{crit} \approx 5.6 \times 10^9 M_{\sun}$ Mpc$^{-3}$ 
(Olive, Steigman, \& Walker 2000).
The metal production by such an intense burst of star formation at $z = 8$
would deplete the hydrogen mass fraction by $\Delta X = 0.1 \pm 0.01$, which exceeds the solar 
value of $\Delta X(\sun) \approx 0.06$. The star formation rate would be lower if only 
high mass stars were formed, but the metal production is still directly linked to the luminosity
density.

Instead of implying interesting phenomena at high redshifts, it is more likely that
the 4.9 $\micron$ rise indicates an incomplete 
subtraction of the Galactic or zodiacal foregrounds (Fig. 7b). 
A Galactic origin for the rise seems unlikely however, since
it would require the 3.5 and 4.9 $\micron$ emission of the ISM to have been underestimated
by a factor of 10 or more, or it would require the presence of some unknown discrete sources 
with spectra that peak at $\sim 5 - 10$ $\micron$ ($T \approx 300 - 600$ K) 
and a distribution that appears isotropic as viewed from the Galactic location of the sun. 
The rise is more likely caused by unsubtracted zodiacal light. 
The zodiacal light spectrum is actually a suspiciously good match to
the residual emission reported by Hauser et al. (1998) from 1.25 -- 100 $\micron$, and to the 
near--IR CIB spectrum reported by others. 
The elevated 4.9 $\micron$ emission should therefore serve
as a warning that residual zodiacal light emission may also affect estimates of the CIB at shorter 
wavelengths. 
The magnitude of this effect is difficult to estimate because the zodiacal light is thermal emission 
at 4.9 $\micron$ and scattered light at shorter wavelengths. Uncertainties in the spatial 
distribution of the dust may produce correlated errors at 4.9 $\micron$ and
shorter wavelengths, whereas uncertainties in the optical properties of the grains may not.

Recent measurements of the optical extragalactic background light (EBL) 
by Bernstein, Freedman \& Madore (2002) indicate that 
high estimates of the 1.25 $\micron$ CIB (e.g. Matsumoto et al. 2000; Cambr\'esy et al. 2001) 
may be affected by systematic errors. At face value, the results would require an unphysically 
sharp rise in the CIB spectrum from 0.8 to 1.1$\micron$. An error in the 1.25 $\micron$ CIB would 
only alter our estimate of the 4.9 $\micron$ CIB by 1--2 nW m$^{-2}$ sr$^{-1}$, but this apparent 
discrepancy provides additional hints at the likelihood and size of systematic errors in the 
analysis of the near--IR CIB and/or the optical EBL. 

In the near future, lower limits on the CIB that will be obtained through deep galaxy counts 
with the Infrared Array Camera (IRAC) on the {\it Space Infrared Telescope Facility} 
({\it SIRTF}) at 3.6, 4.5, 5.8, and 8 $\micron$ should clearly reveal whether or not the 
cosmic star formation rate rises sharply at high $z$ (Fazio et al. 1998).

\section{Near--IR Emission of the Interstellar Medium}

Studies of the relatively faint emission from the diffuse Galactic ISM at 3.5 
and 4.9 $\micron$ require an accurate method of separating it from the 
brighter stellar emission.
Freudenreich (1996) shows that emission of the ISM can be traced to 
high latitudes using the 3.5/2.2 $\micron$ colors, which are relatively 
constant for the stellar population, but are redder where ISM emission 
is present. 
Arendt et al. (1998) correlated the variations in the reddening-free 
near-IR colors with 100 $\micron$ ISM emission to derive the 3.5/100 
$\micron$ and the 4.9/100 $\micron$ colors of the ISM. 
However, the derivation of these colors was restricted to very 
low galactic latitudes. Now, with the models developed here,
we can subtract the stellar emission directly, producing a map of the 3.5 
$\micron$ ISM emission instead of only an average near-IR color. The 3.5 $\micron$
ISM emission derived from subtracting the stellar emission of Model 2
is shown in Figure 8. Comparison with the 100 $\micron$ emission reveals
an excellent match of features down to the effective noise level of the 3.5 $\micron$ 
map. The linear correlation of the 100 and 3.5 $\micron$ emission is plotted
in Figure 9 for regions $|b| < 20\arcdeg$. 
The line plotted over the data is not a direct fit to this correlation, but rather
the expected relation between the 3.5 and 100 $\micron$ emission implied
by the coefficients of Model 2. 
The inset shows that this same linear trend fits all the 
way down to the faintest ISM emission, despite the fact that the faint emission 
has relatively little weight in determining the model coefficients.
Higher latitude emission is all faint, 
and adds noise to the correlation at low intensities, but it still appears to 
follow the same correlation. 

The coefficients $C'$ listed in Table 1 for Models 1 and 2, must be converted to the
actual ISM color through the relation
\begin{equation}
I_{ISM}(3.5)/I_{ISM}(100) = C' + B' I_{ISM}(4.9)/I_{ISM}(100)
\end{equation}
because the observed 4.9 $\micron$ emission, used as a stellar template, 
includes weak ISM emission as well as starlight (see \S2.2). 
ISM emission at 2.2 $\micron$ has not been detected and thus in not 
included in this conversion.
The near-IR colors of the ISM emission determined 
by Arendt et al. (1998) are consistent with this relation for the coefficients 
of either Model 1 or Model 2.

The strong correlation between 3.5 and 100 $\micron$ ISM emission at 
low latitudes as well as high latitudes indicates that, 
in the diffuse ISM, the distribution 
of the very small grains or PAHs responsible for the 3.5 $\micron$ emission 
is not very different from that of the larger grains 
which produce most of the 100 $\micron$ emission.
Previous studies by Giard et al. (1989) and Tanaka et al. (1996) have 
identified similarly good correlation between the 3.3 $\micron$ PAH feature
and 12 and 100 $\micron$ emission using data sets that were less sensitive
or more restricted in coverage than the DIRBE data.
The fact that the correlation is so tight despite Galactic gradients in the 
interstellar radiation field (ISRF), which varies in intensity by factors of $\sim$3 -- 5
on a scale of several kpc (Mathis, Mezger, \& Panagia 1983; Sodroski et al. 1997), 
may be explained by considering the 
heating of the dust grains. First, the 3.5 $\micron$ intensity of the 
small grains or PAHs will be proportional to the UV intensity of the ISRF, while 
the spectral shape of the emission is essentially constant. Second, the 
emission of large grains immersed in the typical ISRF peaks at wavelengths 
slightly longer than 100 $\micron$. Being near the peak of the Planck function,
there is a wide range (a factor of $\gtrsim10^2$) of ISRF intensity, $I_{ISRF}$, 
over which the quantity $I_{ISM}(100) / I_{ISRF}$ varies by no more than
a factor of 2.
Therefore, over a wide range of ISRF intensity, both the 3.5 and 100 $\micron$ emission 
are roughly proportional to the intensity of the ISRF.
Detailed modeling of the emission by Li \& Draine
(2001) shows that the ratio of $I_{ISM}(3.5)/I_{ISM}(100)$ is expected to vary by $<20\%$ 
for an ISRF scaled from 1 -- 30 times its local intensity (see their
Fig. 13). Because the large--scale variation of the ISRF is only a fraction of this 
range, the ratios $I_{ISM}(3.5)/I_{ISM}(60)$ and $I_{ISM}(3.5)/I_{ISM}(240)$ are also observed
to be relatively constant, and correlations of $I_{ISM}(3.5)$ with 
$I_{ISM}(60)$ or $I_{ISM}(240)$ are similar in appearance to Figure 9.
\\
\\

\section{Summary}

We have developed an empirical model of the 3.5 $\micron$ Galactic
and extragalactic emission seen by DIRBE. The Galactic stellar emission is modeled 
with two spatial components. The Galactic ISM emission is modeled with a single spatial template.
The residual emission after subtraction of the Galactic emission is isotropic to a high degree,
but without a significantly positive intensity. However, because our model uses the 
observed emission in other IR bands as spatial templates, 
the residual should be a linear combination 
of the CIB at several wavelengths. Existing estimates of the CIB at 
1.25, 2.2, 3.5, and 100 $\micron$ can be combined to derive a new (but highly uncertain) 
estimate of the CIB intensity at 4.9 $\micron$, which suggests that the 
CIB intensity may begin to increase
at wavelengths $> $3.5 $\micron$. Such a rise in the spectrum of the residual 
emission after subtraction of the foregrounds may indicate interesting physics 
in the spectrum of the CIB, but is more likely to simply indicate systematic errors in the 
subtraction of the zodiacal and/or Galactic foreground emission.

\acknowledgments
We thank the anonymous referee for a careful and helpful review of this paper.
The National Aeronautics and Space Administration/Goddard Space Flight Center 
(NASA/GSFC) was responsible for the design, development, and operation of the {\it COBE}.
GSFC was also responsible for the development of the analysis software and the production
of the mission data sets.
This work was supported by the NASA ADP program, grant NASW-99228.


\begin{deluxetable}{lccccc}
\tabletypesize{\scriptsize}
\tablewidth{0pt}
\tablecaption{Parameters of the 3.5 \lowercase{$\mu {\rm m}$} Model}
\tablehead{
\colhead{Model Version} &
\colhead{$A'$} &
\colhead{$B'$} &
\colhead{$C/C'$~\tablenotemark{a}} &
\colhead{$D'$}
}
\startdata
Dwek \& Arendt & $0.496$ & \nodata & $0.00183$ & \nodata \\
Model 1 & $0.445 \pm 3\times 10^{-4}$ & $0.220 \pm 0.001$ 
 & $0.00135 \pm 7\times 10^{-6}$ & $-0.00197 \pm 6\times 10^{-5}$\\
Model 2 & $0.467 \pm 5\times 10^{-4}$ & $0.248 \pm 0.002$ 
 & $0.00123 \pm 8\times 10^{-6}$ & $-0.00310 \pm 1\times 10^{-4}$\\
\enddata
\tablecomments{Values are for $I_{\nu}$ in units of MJy sr$^{-1}$, with $\pm 1\sigma$
statistical uncertainties of the parameters.}
\tablenotetext{a}{$C$ applies for the Dwek \& Arendt model. $C'$ applies for 
Model 1 and Model 2.}
\end{deluxetable}

\begin{deluxetable}{lccccc}
\tabletypesize{\scriptsize}
\tablewidth{0pt}
\tablecaption{Systematic Uncertainties of the Mean 3.5 \lowercase{$\mu {\rm m}$}
Residual Emission}
\tablehead{
\colhead{} &
\multicolumn{2}{c}{Model 1} & &
\multicolumn{2}{c}{Model 2} \\
\cline{2-3}\cline{5-6}
\colhead{Uncertainty} &
\colhead{$\sigma$ (nW m$^{-2}$ sr$^{-1}$)} & 
\colhead{Scale Factor} & &
\colhead{$\sigma$ (nW m$^{-2}$ sr$^{-1}$)} & 
\colhead{Scale Factor}
}
\startdata
Zodiacal Light (1.25 $\micron$) & \nodata & \nodata & & 15 & 0.0368 \\
Zodiacal Light (2.2 $\micron$) & 6 & 0.280 & & 6 & 0.355 \\
Zodiacal Light (3.5 $\micron$) & 2 & 1.00 & & 2 & 1.00 \\
Zodiacal Light (4.9 $\micron$) & 6 & 0.308 & & 6 & 0.360 \\
Zodiacal Light (100 $\micron$) & 6 & 0.0386 & & 6 & 0.0351 \\
Stellar \& ISM Model          & 2.2 & 1.00 & & 1.7 & 1.00 \\
Total                          & 3.9 & \nodata & & 4.1 & \nodata \\
\enddata
\end{deluxetable}

\begin{deluxetable}{lccccc}
\tabletypesize{\scriptsize}
\tablewidth{0pt}
\tablecaption{Mean 3.5 \lowercase{$\mu {\rm m}$} Residual Emission at Select Patches}
\tablehead{
\colhead{Patch} & 
\colhead{Size/Location} &
\colhead{Hauser et al. (1998)} &
\colhead{Dwek \& Arendt (1998)} &
\colhead{Model 1} &
\colhead{Model 2}
}
\startdata
NEP  & $10\arcdeg\times 10\arcdeg$ at $\beta = +90$ & $ 5.6$ & $ 5.8$ & $-2.3$ & $-1.8$\\
SEP  & $10\arcdeg\times 10\arcdeg$ at $\beta = -90$ & $-2.3$ & $ 6.6$ & $-1.8$ & $-1.4$\\
NGP  & $10\arcdeg\times 10\arcdeg$ at $b = +90$ & $15.2$ & $11.4$ & $-0.9$ & $-2.2$\\
SGP  & $10\arcdeg\times 10\arcdeg$ at $b = -90$ & $15.9$ & $12.7$ & $-1.1$ & $-2.7$\\
LH   & $ 5\arcdeg\times  5\arcdeg$ at $(l,b) = (150\arcdeg,+53\arcdeg)$ & $16.1$ & $11.0$ & $-0.1$ & $-1.0$\\
HQA  & $|b| > 30$ and $|\beta| > 25$ & $10.5$ & $ 9.9$ & $-0.8$ & $-1.5$\\
HQAN & $ b  >+30$ and $ \beta  >+25$ & $11.3$ & $ 9.5$ & $-0.9$ & $-1.5$\\
HQAS & $|b| <-30$ and $ \beta  <-25$ & $ 9.7$ & $10.2$ & $-0.8$ & $-1.6$\\
HQB  & $|b| > 60$ and $|\beta| > 45$ & $11.4$ & $ 9.9$ & $-0.4$ & $-1.1$\\
HQBN & $ b  >+60$ and $ \beta  >+45$ & $11.7$ & $ 9.5$ & $-0.5$ & $-1.1$\\
HQBS & $ b  <-60$ and $ \beta  <-45$ & $11.0$ & $10.2$ & $-0.3$ & $-1.2$\\
\enddata
\tablecomments{Results are for $\nu I_{\nu}$ in units of nW m$^{-2}$ sr$^{-1}$.}
\end{deluxetable}

\begin{deluxetable}{ccccc}
\tabletypesize{\scriptsize}
\tablewidth{0pt}
\tablecaption{Gaussian Fits to 3.5 \lowercase{$\mu {\rm m}$} Residual Emission Histograms}
\tablehead{
\colhead{Model Version} &
\colhead{Region} &
\colhead{Mean $\nu I_{\nu}$} &
\colhead{$\sigma(\nu I_{\nu})$} &
\colhead{K-S Probability} \\
\colhead{} &
\colhead{} &
\colhead{(nW m$^{-2}$ sr$^{-1}$)} &
\colhead{(nW m$^{-2}$ sr$^{-1}$)} &
\colhead{Non-Gaussian~\tablenotemark{a}}
}
\startdata
Hauser et al. (1998)  & HQAN & $ 5.70$ & $6.89$ & $1.00$ \\
Hauser et al. (1998)  & HQAS & $ 4.25$ & $8.50$ & $1.00$ \\
Dwek \& Arendt (1998) & HQAN & $ 9.35$ & $2.23$ & $1.00$ \\
Dwek \& Arendt (1998) & HQAS & $10.09$ & $2.43$ & $1.00$ \\
Model 1               & HQAN & $-0.86$ & $1.77$ & $1.00$ \\
Model 1               & HQAS & $-0.83$ & $1.91$ & $1.00$ \\
Model 2               & HQAN & $-1.42$ & $1.85$ & $1.00$ \\
Model 2               & HQAS & $-1.54$ & $2.11$ & $1.00$ \\
Hauser et al. (1998)  & HQBN & $ 5.05$ & $3.93$ & $1.00$ \\
Hauser et al. (1998)  & HQBS & $ 3.96$ & $5.47$ & $1.00$ \\
Dwek \& Arendt (1998) & HQBN & $ 9.44$ & $1.48$ & $0.93$ \\
Dwek \& Arendt (1998) & HQBS & $10.19$ & $1.74$ & $0.92$ \\
Model 1               & HQBN & $-0.48$ & $1.44$ & $0.85$ \\
Model 1               & HQBS & $-0.39$ & $1.64$ & $0.98$ \\
Model 2               & HQBN & $-1.12$ & $1.51$ & $0.48$ \\
Model 2               & HQBS & $-1.21$ & $1.71$ & $0.96$ \\
\enddata
\tablenotetext{a}{
1.00 implies a distinctly non-gaussian distribution. 0.00 implies
that the observed distribution is perfectly Gaussian. 0.50 implies a 
distribution is indistinguishable
from Gaussian given the expected random sampling variance.}
\end{deluxetable}

\begin{deluxetable}{cccc}
\tablewidth{0pt}
\tablecaption{Galactic Gradients of the 3.5 \lowercase{$\mu {\rm m}$} Residual Emission}
\tablehead{
\colhead{Model Version} &
\colhead{Region} &
\colhead{Gradient} &
\colhead{Correlation}\\
\colhead{} &
\colhead{} &
\colhead{(nW m$^{-2}$ sr$^{-1}$)/$\csc(|b|)$} &
\colhead{Coefficient}
}
\startdata
Hauser et al. (1998)  & HQA & $ -9.91 \pm 0.23$ & $-0.16$ \\
Dwek \& Arendt (1998) & HQA & $ -3.32 \pm 0.03$ & $-0.34$ \\
Model 1               & HQA & $ -0.95 \pm 0.03$ & $-0.12$ \\
Model 2               & HQA & $  0.35 \pm 0.03$ & $ 0.04$ \\
Hauser et al. (1998)  & HQB & $-19.45 \pm 5.32$ & $-0.04$ \\
Dwek \& Arendt (1998) & HQB & $ -8.61 \pm 0.70$ & $-0.14$ \\
Model 1               & HQB & $ -2.53 \pm 0.66$ & $-0.05$ \\
Model 2               & HQB & $  1.31 \pm 0.67$ & $ 0.02$ \\
\enddata
\end{deluxetable}

\begin{deluxetable}{cccccc}
\tabletypesize{\scriptsize}
\tablewidth{0pt}
\tablecaption{Implied 4.9 \lowercase{$\mu {\rm m}$} CIB Intensity}
\tablehead{
\colhead{}&
\multicolumn{2}{c}{Wright et al. Zodiacal Light Model}&
\colhead{}&
\multicolumn{2}{c}{Kelsall et al. Zodiacal Light Model}\\
\cline{2-3}\cline{5-6}\\
\colhead{} &
\colhead{Intensity} &
\colhead{} &
\colhead{}&
\colhead{Intensity} &
\colhead{}\\
\colhead{} &
\colhead{(nW m$^{-2}$ sr$^{-1}$)} &
\colhead{Reference} &
\colhead{}&
\colhead{(nW m$^{-2}$ sr$^{-1}$)} &
\colhead{Reference}
}
\startdata
$\nu I_{CIB}(1.25)$ & 29 $\pm$ 16 & Wright (2001) & & 54 $\pm$ 17 & Cambr\'esy et al. (2001)\\
$\nu I_{CIB}(2.2)$ & 20 $\pm$ 6 & Wright (2001) & & 28 $\pm$ 7 & Cambr\'esy et al. (2001)\\
$\nu I_{CIB}(3.5)$ & 12 $\pm$ 3 & Wright \& Reese (2000) & & 16 $\pm$ 4 & Dwek \& Arendt (1998)\tablenotemark{a}\\
$\nu I_{CIB}(100)$ & 23 $\pm$ 6 &Lagache et al. (2000)\tablenotemark{b} & & 23 $\pm$ 6 & Lagache et al. (2000)\\
\hline
$\nu I_{CIB}(4.9)$ [Model 1] & 23 $\pm$ 13 & & & 29 $\pm$ 16 & \\
$\nu I_{CIB}(4.9)$ [Model 2] & 22 $\pm$ 12 & & & 27 $\pm$ 15 & \\
\enddata
\tablenotetext{a}{Based on $\nu I_{CIB}(2.2)$ = 28 $\pm$ 7 nW m$^{-2}$ sr$^{-1}$.}
\tablenotetext{b}{The Lagache et al. (2000) estimate is used here, even though it was derived 
with a Kelsall et al. zodiacal light model.}
\end{deluxetable}

\newpage

\begin{figure}
\caption{Images of 3.5 $\micron$ residual emission after subtraction of
foreground emission models. (a) Hauser et al. (1998), (b) Dwek \& 
Arendt (1998), (c) Model 1, (d) Model 2. The images are shown as the Galactic 
azimuthal equal-area projections used by Hauser et al. with the north 
Galactic hemisphere on the left and south on the right. The Galactic 
center is at the bottom edge of each hemispherical image. 
The intensity ranges are scaled linearly over the range $\pm 0.03$ 
MJy sr$^{-1}$ about the median value of each image (median values = 
0.0076, 0.0125, -0.0020, -0.0030 MJy sr$^{-1}$ respectively). 
The bright source blanking that was applied to
the Hauser et al. data was also applied to the other images.
This causes the small holes at the locations of bright stars, and large
holes at the locations of the Small and Large Magellanic Clouds.}
\end{figure}
\setcounter{figure}{0}
\begin{figure}
\caption{{\it Continued.}}
\end{figure}

\begin{figure}
\caption{Histograms of the residual 3.5 $\micron$ intensities in HQAN and 
HQAS regions from (a,b) the Hauser et al. (1998) results, (c,d) the Dwek
\& Arendt (1998) results,(e,f) Model 1, and (g, h) Model 2. The smooth lines in 
panels (c) -- (h) show Gaussian fits to the histograms. The histograms
in (a) and (b) are fit with an additional quadratic base level, which 
mitigates the effect of the positive tail on the Gaussian fit to the peak.
The HQAN results are on the left, and HQAS results are on the right.}
\end{figure}

\begin{figure}
\caption{Same as Figure 2, but for the smaller HQBN and HQBS regions.}
\end{figure}

\begin{figure}
\caption{Gradients of the residual 3.5 $\micron$ emission with respect
to the cosecant of the Galactic latitude, $\csc{b}$. Panels (a,b) 
show the Hauser et al. (1998) results, (c,d) show the Dwek et al. (1998) 
results, (e,f) and (g,h) show the results of Model 1 and Model 2. 
($\csc{15\arcdeg} \approx 4$, $\csc{30\arcdeg} = 2.0$, and 
$\csc{60\arcdeg} = 1.15$).}
\end{figure}

\begin{figure}
\caption{Two-point correlation functions of the 3.5 $\micron$ residual emission 
calculated in the HQB regions for (a) the Hauser et al. (1998) result, (b) 
the Dwek \& Arendt (1998) result, (c) the Model 1 result, and (d) the Model 2 
result. The solid lines in each panel are the estimated $\pm 1\sigma$ 
uncertainties. Note the change in scale between the Hauser et al. results and the 
others.}
\end{figure}

\begin{figure}
\caption{This shows the excess residual emission at 4.9 $\micron$ 
after removal of zodiacal and Galactic foregrounds.
The upper curve (circles) shows 1.25 -- 3.5 $\micron$ CIB estimates 
($\pm 1 \sigma$) made using 
the Kelsall et al. (1998) model for subtraction of zodiacal light.
The 4.9 $\micron$ data point is extrapolated from these 
prior results using eq.~(16). (See Table 6.)
The lower curve (squares) shows the same for results using the 
zodiacal light model of Wright (2001). 
The dashed lines are power--law extrapolations of 
the 1.25 -- 3.5 $\micron$ data points. 
Diamonds indicated the residual emission levels measured by Hauser et al. 
(1998) and {\it not} identified as CIB because of the large 
uncertainties and lack of isotropy.}
\end{figure}

\begin{figure}
\epsscale{1.00}
\caption{(a) This figure compares several model galaxy spectra 
with the residual IR emission. 
Assumed CIB intensities at 1.25, 2.2, and 
3.5 $\micron$ (using the Kelsall et al. zodiacal light model, as in 
Fig. 6), 
and the implied CIB at 4.9 $\micron$ 
are plotted as circles (with 1 $\sigma$ error bars). 
The residual emission at all DIRBE wavelengths found by Hauser 
et al. (1998) are shown as diamonds.
Crosses indicate the optical EBL measured by Bernstein et al. (2002). 
Chary \& Elbaz (2001) model galaxy spectra at $L_{gal} = 
1.1\times 10^{11}$ and $3.8\times 10^{13} L_{\sun}$, 
normalized at 2.2 $\micron$, are shown as dashed and dotted lines. 
The solid line shows a $1.8\times 10^9 L_{\sun}$ spectrum redshifted to 
$z = 8$, and normalized to fit the 3.5 and 4.9 $\micron$ residuals.
(b) This shows the same residuals as above, but compared to possible 
local sources of systematic errors: a 
high--latitude zodiacal spectrum (scaled by $\sim 1/7$; dashed line), 
a high--latitude ISM spectrum (scaled by $\sim 20$; solid line), 
and a high--latitude Galactic stellar spectrum 
(scaled by $\sim 1/5$; dotted line).}
\end{figure}

\begin{figure}
\caption{The top panel shows the 100 $\micron$ ISM emission on a 
logarithmic intensity scale (0.1 -- 10$^{4}$ MJy/sr) and a 
Mollweide projection. The lower panel shows the 3.5 $\micron$ ISM 
emission on a logarithmic scale (0.1 -- 10$^{4}$ kJy/sr), 
after subtraction of the Model 2 stellar emission.
No bright source blanking has been applied.}
\end{figure}

\begin{figure}
\caption{ISM emission at 100 and 3.5 $\micron$ exhibits a tight linear
correlation across a wide range of intensities. The inset is an expanded 
view of the plot at low intensities. The intensities plotted
here are from pixels at $|b| < 20\arcdeg$. Including higher latitude pixels
would add more data at the low intensity end of the plots, but would
not alter the trends shown. The slope of the line drawn in the 
main panel and inset is the parameter $C$ from Model 2 (Table 1).}
\end{figure}


\begin{references}
\reference{}{Arendt, R. G., et al. 1994, \apj, 425, L85}
\reference{}{Arendt, R. G., et al. 1998, \apj, 508, 74}
\reference{}{Bernard, J. P., Boulanger, F., D\'esert, F. X., Giard, M., Helou, G.,
\& Puget, J. L. 1994, \aap, 291, L5}
\reference{}{Bernstein, R. A., Freedman, W. L, \& Madore, B. F. 2002, \apj, 571, 56}
\reference{}{Boggess, N., et al. 1992, \apj, 397, 420}
\reference{}{Cambr\'esy, L., Reach, W. T., Beichman, C. A. \& Jarrett, T. H. 2001, \apj, 555, 563}
\reference{}{Chary, R. R., \& Elbaz, D. 2001, \apj, 556, 562}
\reference{}{Dwek, E., \& Arendt, R. G. 1998, \apj, 508, L9}
\reference{}{Fazio, G., et al. 1998, Infrared Astronomical Instrumentation, \procspie, 3354,1024}
\reference{}{Finkbeiner, D. P., Davis, M., \& Schlegel, D. J. 2000, \apj, 544, 81}
\reference{}{Freudenreich, H. T. 1996, \apj, 468, 663}
\reference{}{Giard, M., Pajot, F., Lamarre, J. M., Serra, G. , \& Caux, E. 1989, \aap, 215, 92}
\reference{}{Gorjian, V., Wright, E. L., \& Chary, R. R. 2000, \apj, 536, 550}
\reference{} {Hauser, M. G., Kelsall, T., Leisawitz, D., \& Weiland, J. eds. 
1997, {\it COBE Diffuse Infrared Background Experiment (DIRBE)
Explanatory Supplement}, COBE Ref. Pub. No.  97-A (Greenbelt, MD:
NASA/GSFC), available in electronic form from the NSSDC}
\reference{}{Hauser, M. G., et al. 1998, \apj, 508, 25}
\reference{}{Hauser, M. G., \& Dwek, E. 2001, \araa, 39, 249}
\reference{}{Kashlinsky, A., Mather, J. C., Odenwald, S., \& Hauser, M. G. 1996a, \apj, 470, 681}
\reference{}{Kashlinsky, A., Mather, J. C., \& Odenwald, S. 1996b, \apj, 473, L9}
\reference{}{Kashlinsky, A., \& Odenwald, S. 2000, \apj, 528, 74}
\reference{}{Kelsall, T., et al. 1998, \apj, 508, 44}
\reference{}{Lagache, G., Haffner, L. M., Reynolds, R. J., \& Tufte, S. L. 2000, \aap, 354, 247}
\reference{}{Lanzetta, K. M., Yahata, N., Pascarelle, S., Chen, H.-W., \& Fern\'andez-Soto, A.
2001, preprint (astro--ph/011129)}
\reference{}{Li, A., \& Draine, B. T. 2001, \apj, 554, 778}
\reference{}{Mathis, J. S., Mezger, P. G., \& Panagia, N. 1983, \aap, 128, 212}
\reference{}{Matsumoto, T., et al. 2000, ISO Surveys of a Dusty Universe, 
Lemke, D, Stickel, M., \& Wilke, K., eds. (Berlin: Springer--Verlag), 96}
\reference{}{Odenwald, S., Newmark, J., \& Smoot, G. 1998, \apj, 500, 554}
\reference{}{Olive, K. A., Steigman, G., \& Walker, T. P. 2000, \physrep, 333/334, 389}
\reference{}{Rieke, G. H., \& Lebofsky, M. J. 1985, \apj, 288, 618}
\reference{}{Siverberg, R. F., et al. 1993, \procspie, 2019, 180}
\reference{}{Sodroski, T. J., Odegard, N., Arendt, R. G., Dwek, E., Weiland, J. L., 
Hauser, M. G., \& Kelsall, T. 1997, \apj, 480, 173}
\reference{}{Tanaka, M., Matsumoto, T., Murakami, H., Kawada, M., Noda, M., 
\& Matsuura, S. 1996, \pasj, 48, L53}
\reference{}{Wainscoat, R. J., Cohen, M., Volk, K., Walker, H. J., \& Schwartz, D. E. 1992, 
\apjs, 83, 111}
\reference{}{Wright, E. L. 2001, \apj, 553, 538}
\reference{}{Wright, E. L., \& Reese, E. D. 2000, \apj, 545, 43}
\end{references}
\end{document}